## Public Battery Field Data

Robert Masse
Astrolabe Analytics, Inc.

### Abstract

Battery systems and algorithms are developed and evaluated primarily on laboratory cycling data, but fielded battery systems are subjected to duty cycles, temperatures, and measurement conditions that cannot be reproduced in a lab. Field data from batteries in service has begun to appear in public releases attached to individual papers, but each release has its own format, units, labels, and other gaps that make it difficult to compare against others.

We present a curated collection of 35 public field-data releases from 27 research groups, covering applications that span passenger EVs, buses, light electric mobility, grid-scale storage, distributed storage, consumer electronics, industrial robotics and satellites. Where the releases are complete enough to measure, this heterogeneous collection contains 256 GB of data from at least:

- 149,188 lithium-ion cells,
- 3,187 batteries,
- 115.6 MWh of energy storage capacity, and
- 3,040 unit-years of service.

Critically, it also contains 6,465 labeled or reported faults, which provide greater visibility into what real battery performance and safety issues look like outside the lab.

We release a public repository of metadata and code for each of the field datasets. For each dataset it provides a registry entry with sources, license and access links, checksums to confirm a download is complete, a metadata table, and code for how to load it. We also report where released datasets differ from their papers.

We hope the collection can help serve researchers (human or otherwise) as a useful platform for improving battery field diagnostics, fault detection and health estimation. Better field evidence can also feed back to improve manufacturing quality, inform second-life screening and residual-value estimates, and broadly accelerate electrification.

### 1. Introduction

This collection was assembled to help answer questions about battery failure and degradation in the field. Whereas laboratory cycling data can help understand degradation under controlled conditions, fielded battery systems face a more complicated set of stressors. These include the end-user's duty cycle, environmental conditions (temperature, water ingress), and faults that arise when individual cells are assembled into packs and modules. These issues can manifest as individual weak cells, faults in the electronics, or other downstream effects from a wider system malfunction. In fact, in grid-scale storage only 3 of 26 failures with a known cause (12%) were traced to the cells.[1] The rest were from the control systems (battery and energy management) or in the surrounding equipment, such as cabling, power conversion, cooling and fire suppression.

In recent years, a large volume of laboratory data has been published as part of a positive trend towards open science.[2,3] Field data has followed more slowly, but nevertheless the pace has picked up, with 18 of the 35 releases in this collection appearing in 2025 and 2026 alone (Figure 1).

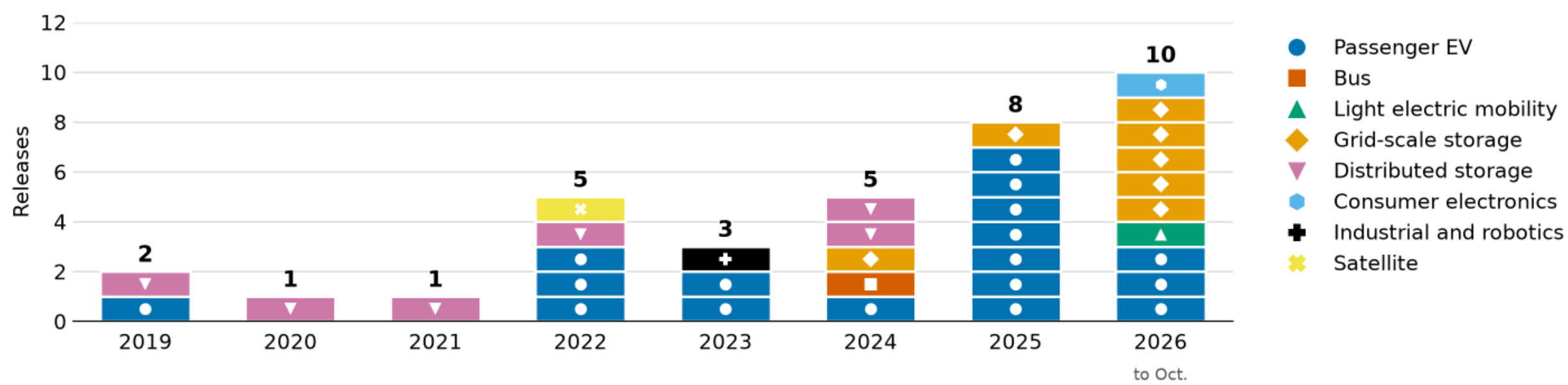

**Figure 1.** Public field-data releases in the collection by year first published online (paper, or data record where there is no paper), as of October 2026.

Taken one at a time, these datasets have value for studying battery aging in specific contexts. However, they can be difficult to work with. They come in different data formats (parquet, CSV, pickle, RAR) with different schemas for what data was recorded, and no common convention for identifiers, labels, or definitions for what a given data field means. In some cases, there appear to be discrepancies between what was released and what the authors claimed to release in the paper (Section 2.2).

Taken together, this collection presents an opportunity to compare commercially relevant batteries across applications and chemistries, using data no single organization could otherwise assemble. They contain records of batteries in use for more than eight years, through seasons, rests and changes in how the owner operated them.[4,5] They also contain faults as the authors labeled or reported them. Examples include cells with micro-short circuits in a light electric mobility fleet,[6] faulty and abnormal vehicles in passenger EV fleets,[7–9] systems returned under warranty,[10] solar home batteries that went to repair for capacity loss,[29] faulty charging sessions at public stations,[36] and a cell bypassed in a 300-cell lead-acid string after years of grid service.[4] Section 2.3 gives the counts. In some releases the battery management system (BMS) health estimate can be compared against an independently measured reference, such as annual full-power tests on a utility-scale storage system[5] and capacity tests on home storage systems.[11]

The main purpose of the collection is to serve as a foundation for working with available field data and surfacing previously intractable insights. We do not present analysis results here beyond what is needed to check that the data match what their papers describe (unit and cell counts, time spans, energy and fault labels). Further analysis belongs to the papers that use this collection, many of which we believe are waiting to be written.

## 2. Data Description

### 2.1. Inclusion Criteria

There are now over 300 publicly available battery datasets available from repositories such as Zenodo, figshare, and Mendeley Data.[3] Here we limited ourselves to releases where the battery was used outside of a laboratory: driven on the road, cycled on the grid, and so on. Where a release also includes reference performance tests on the same batteries, that data is also included as "ground truth." To make this a more manageable task, laboratory-aged cells are not considered here, nor are simulated or synthetic datasets. Because the criterion pertains to how the batteries are used, we also include four datasets featuring lead-acid batteries.[4,12,20,29]

We also only include datasets that are available without a subscription. For this reason two subscription-only datasets on IEEE DataPort were excluded.[13,14] Candidates were identified and believed to be fully accounted for as of October 2026.

### 2.2. Data Access and Contents

Table 1 lists each dataset in order of the year it was published, with links to its paper, data record, code and its license. Table 2 lists each package's application, country, chemistry, units, cells, service time, and energy (when it is possible to observe or calculate the value). Section 3.5 describes how energy was obtained for each release. In four releases the number of units or cells differs from the paper. Table 2 gives the release value, which is used in every total, with the paper's in parentheses.

Together, the releases cover 3,187 batteries. Of these, 1,939 are in vehicles (1,653 passenger EVs, 271 light electric mobility vehicles, 10 industrial vehicles and 5 buses). Another 1,211 are stationary storage systems, 33 are consumer devices and 4 are satellite batteries.[33] Their energy totals about 115.6 MWh. Releases with timestamps give 3,040 unit-years of service, and releases without timestamps add at least 331,000 observation-hours (37.8 unit-years). One release of charging records has no vehicle identifiers.[36] Its 31,449 charging sessions are kept as a separate group and are not counted as unique batteries or in energy, but their observation-hours are included in the 331,000.

The code described in Section 3.3 reads each deposit exactly as published, so users download the data from the links in Table 1 and need no intermediate copy.

**Table 1.** Sources of the releases, sorted by year first published online (paper, or data record where there is no paper). Links name the host. Licenses marked * restrict commercial use, and ** state no license.

| Year | Reference | Package Name | Data | Code | License |
|---|---|---|---|---|---|
| 2019 | Vink, Ankyu and Koyama, Scientific Data[12] | tsukuba | figshare | In deposit | CC0 1.0 |
| 2019 | Oh, LeBlanc and Peng, IEEE Trans. Intell. Transp. Syst.[30] | ved | GitHub | GitHub | Apache 2.0 |
| 2020 | Gaura et al., Zenodo[31,42] | brazil_star | Zenodo | None | CC BY 4.0 |
| 2021 | Aitio and Howey, Joule[29] | aitio | Oxford ORA | None | CC BY-NC 4.0 * |
| 2022 | Wei et al., Journal of Energy Storage[32] | beijing_ev | Mendeley Data | None | CC BY 4.0 |
| 2022 | Jara et al., Data in Brief[33] | birds_cubesat | Mendeley Data | None | CC BY 4.0 |
| 2022 | Gaukema et al., Zenodo[15,43] | cloverleaf | Zenodo | None | CC BY 4.0 |
| 2022 | He et al., arXiv[9] | evbattery | figshare | GitHub | CC BY-NC-SA 4.0 * |
| 2022 | Wassiliadis et al., eTransportation[34] | tum_id3_2022 | mediaTUM | None | CC BY 4.0 |
| 2023 | Deng et al., Applied Energy[16] | deng | GitHub | GitHub | MIT (repository) |
| 2023 | Zarfati and Bedogni, IEEE CCNC[17] | flashbattery-agv | Zenodo | None | CC BY 4.0 |
| 2023 | Zhang et al., Nature Communications[8] | zhang2023 | figshare | GitHub | CC BY 4.0 |
| 2024 | Bilfinger et al., eTransportation[18] | bilfinger2024 | mediaTUM | GitHub | CC BY 4.0 |
| 2024 | Fei, Mendeley Data[19] | fei_bus | Mendeley Data | None | CC BY 4.0 |
| 2024 | Evaluation operation report, RWTH Publications[20] | m5bat-2023-04 | RWTH Publications | None | CC BY 4.0 |
| 2024 | Figgener et al., Nature Energy[11] | rwth-home | Zenodo | Zenodo | CC BY 4.0 |
| 2024 | Schaeffer et al., Cell Reports Physical Science[10] | schaeffer | Zenodo | GitHub | CC BY-NC 4.0 * |
| 2025 | Bilfinger et al., npj Clean Energy[41] | bilfinger2025 | mediaTUM | GitHub | CC BY 4.0 |
| 2025 | Cao et al., Nature Communications[7] | cao | Zenodo | Zenodo | CC BY 4.0 |
| 2025 | Liu et al., Nature Communications[21] | changan | Changan portal | GitHub | CC BY-NC-SA 4.0 * |
| 2025 | Yasko et al., Scientific Data[22] | ku_leuven_bev | KU Leuven RDR | GitHub | CC BY 4.0 |
| 2025 | Fast et al., Zenodo[35] | rise_conbat | Zenodo | None | CC BY 4.0 |
| 2025 | Yang et al., Nature Communications[36] | shenzhen_charging | Mendeley Data | None | CC BY 4.0 |
| 2025 | Tao et al., eTransportation[37] | tongji_rest | Zenodo | None | CC BY 4.0 |
| 2025 | Schreiber et al., eTransportation[24] | tumftm | GitHub | GitHub | CC BY-NC 4.0 * |
| 2026 | Bilfinger et al., eTransportation[25] | bilfinger2026 | mediaTUM | GitHub | CC BY 4.0 |
| 2026 | He et al., Zenodo[38] | cell_to_pack | Zenodo | Zenodo | CC BY 4.0 |
| 2026 | Li et al., Cell Reports Physical Science[26] | li2026 | Zenodo | GitHub | CC BY 4.0 |
| 2026 | Zurmühlen, Koltermann and Sauer, Energies[4] | m5bat-pbacid | RWTH Publications | None | CC BY 4.0 |
| 2026 | Crawford and Thompson, Zenodo[39,44] | pnnl_epri | Zenodo | Zenodo | CC BY 4.0 |
| 2026 | Kyeremeh et al., IEEE Access[5] | ppl | GitHub | GitHub | Unstated ** |
| 2026 | Laufen et al., RWTH Publications[23] | rwth-android | RWTH Publications | None | CC BY 4.0 |
| 2026 | Lanubile et al., Stanford Digital Repository[40] | stanford_ev | Stanford Digital Repository | None | CC BY-NC-ND 4.0 * |
| 2026 | Xie et al., Cell Reports Physical Science[6] | xie | Zenodo | Zenodo | CC BY 4.0 (data), MIT (code) |
| 2026 | Zhou et al., Nature Energy[27] | zhou2026 | OneDrive | GitHub | Unstated (data), MIT (code) ** |

**Table 2.** Dataset contents. Service time is unit-years where the release has timestamps, or otherwise states a lower bound on observation-hours (h). Energy basis: Stated is a rating from the paper or record. Computed is from the stated cell configuration or from the released data (Section 3.5). Chemistry: NMC-class is inferred from cell voltage (NMC, NCA and LMO cannot readily be told apart this way). Li-ion means the cathode chemistry is not stated. Lead-acid includes flooded, gel and valve-regulated designs. Where a release's unit or cell count differs from its paper, the release value is shown with the paper's in parentheses. † Some of these units are counted under another release (Table 5). § Energy covers only some of the release's units (Section 3.5).

| Package Name | Application | Country | Chemistry | Units | Cells | Service time | Energy, MWh (basis) | Size, GB |
|---|---|---|---|---|---|---|---|---|
| tsukuba | Distributed storage | Japan | Lead-acid | 1 | Not in release | 3.3 unit-years | 0.326 (stated) | 2.73 |
| ved | Passenger EV | United States | Not stated | 28 | Not in release | 23.9 unit-years | 0.072 (stated) § | 0.18 |
| brazil_star | Distributed storage | Brazil | Not stated | 4 | Not in release | 2.8 unit-years | 0.001 (computed) § | <0.01 |
| aitio | Distributed storage | Sub-Saharan Africa | Lead-acid | 1,027 | 6,162 lead-acid | 1,665.0 unit-years | 0.246 (computed) | 3.04 |
| beijing_ev | Passenger EV | China | Not stated | 1 | Not in release | 1.6 unit-years | 0.032 (computed) | <0.01 |
| birds_cubesat | Satellite | In orbit | NiMH | 4 | Not in release | At least 78 h | <0.001 (computed) | <0.01 |
| cloverleaf | Distributed storage | Belgium | NMC-class | 3 | Not in release | 1.0 unit-years | 0.080 (computed) | <0.01 |
| evbattery | Passenger EV | China | NMC-class | 464 † | Not in release | At least 57,265 h | Not available | 1.41 |
| tum_id3_2022 | Passenger EV | Germany | Not stated | 1 † | 216 Li-ion | <0.1 unit-years | Counted under tumftm | 2.91 |
| deng | Passenger EV | China | NMC | 20 | 1,800 Li-ion | 46.2 unit-years | 0.953 (computed) | 0.23 |
| flashbattery-agv | Industrial and robotics | Italy | Li-ion | 10 | Not in release | 2.1 unit-years | Not available | <0.01 |
| zhang2023 | Passenger EV | China | NMC-class | 347 | Not in release | At least 45,030 h | Not available | 1.48 |
| bilfinger2024 | Passenger EV | Germany | NMC, LFP | 2 | 322 Li-ion | 3.2 unit-years | 0.111 (stated) | 0.30 |
| fei_bus | Bus | Not stated | Not stated | 4 | Not in release | No timestamps | Not available | <0.01 |
| m5bat-2023-04 | Grid-scale storage | Germany | LMO/NMC, LFP, LTO, lead-acid[28] | 10 † | 24,672 Li-ion, 1,222 lead-acid | 0.7 unit-years | 6.714 (stated) | 0.10 |
| rwth-home | Distributed storage | Germany | LFP, LMO/NMC, NMC | 21 | 12,036 Li-ion | 120.1 unit-years | 0.153 (stated) | 21.12 |
| schaeffer | Distributed storage | Not stated | LFP | 28 | 224 Li-ion | 56.9 unit-years | 0.108 (computed) | 1.67 |
| bilfinger2025 | Passenger EV | Germany | NMC, LFP | 12 † | 9,840 Li-ion | 6.9 unit-years | 0.373 (stated) | 0.31 |
| cao | Passenger EV | China | NMC-class, LFP | 514 (paper: 515) | 50,030 Li-ion | At least 151,900 h | 15.107 (computed) § | 38.06 |
| changan | Passenger EV | China | NMC | 300 | 28,800 Li-ion | 975.4 unit-years | 16.294 (computed) | 79.69 |
| ku_leuven_bev | Passenger EV | Belgium | LFP, NCA | 2 | Not in release | 1.8 unit-years | 0.139 (stated) | 0.16 |
| rise_conbat | Grid-scale storage (second-life) | Sweden | Not stated | 88 | 1,056 Li-ion | 56.5 unit-years | 0.115 (stated) | 0.71 |
| shenzhen_charging | Passenger EV | China | LFP, NMC, LMO, LCO, lithium polymer (Section 3.4) | 31,449 charging sessions | Not in release | At least 21,479 h | Not counted | 0.22 |
| tongji_rest | Passenger EV | China | LFP, NMC | 4 | 432 Li-ion | 9.9 unit-years | 0.189 (computed) | 0.13 |
| tumftm | Passenger EV | Germany | NMC | 7 | 1,512 Li-ion | 6.7 unit-years | 0.406 (stated) | 0.86 |
| bilfinger2026 | Passenger EV | Germany | NMC, LFP | 7 † | 1,402 Li-ion | 0.3 unit-years | 0.401 (stated) | 0.30 |
| cell_to_pack | Grid-scale storage | Not stated | LFP, NMC | 16 | 6,656 Li-ion | At least 54,750 h | 6.500 (computed) | 3.54 |
| li2026 | Grid-scale storage | Not stated | LFP | 1 | 240 Li-ion (paper: 200) | 1,203 h | 0.156 (computed) | 0.02 |
| m5bat-pbacid | Grid-scale storage | Germany | Lead-acid | 1 | 300 lead-acid | 8.1 unit-years | 1.066 (stated) | 4.65 |
| pnnl_epri | Grid-scale storage | United States | LFP, NMC, LMO/NMC, | 12 † | Not in release | 17.2 unit-years | 63.870 (stated and computed) § | 3.43 |

| Package Name | Application | Country | Chemistry | Units | Cells | Service time | Energy, MWh (basis) | Size, GB |
|---|---|---|---|---|---|---|---|---|
| | | | vanadium flow | | | | | |
| ppl | Grid-scale storage | United States | Li-ion | 1 | 4,760 Li-ion | 8.3 unit-years | 2.000 (stated) | 67.59 |
| rwth-android | Consumer electronics | Germany | Li-ion | 33 | 33 Li-ion | 3.5 unit-years | 0.001 (computed) | 0.08 |
| stanford_ev | Passenger EV | United States | Not stated | 1 | 108 Li-ion | 3.5 unit-years | 0.091 (computed) | 18.30 |
| xie | Light electric mobility | Not stated | LFP | 271 | 5,241 Li-ion (paper: 5,234) | 4.8 unit-years | Not available | 0.16 |
| zhou2026 | Passenger EV, bus | China | NMC, LFP | 3 (paper: 133) | 346 Li-ion | 10.4 unit-years | 0.209 (stated) | 2.15 |

### 2.3. Labeled and Reported Faults

One of the most notable features of this collection is that it contains 6,465 labeled or reported faults (Table 3). Of these, 6,435 come from labels released with the data and 30 are reported by the authors. Each paper defines a fault in its own way, and the definition is not always clear.

Released labels include per-cell labels for micro-short circuits, low capacity and abnormally high or low state of charge in a light electric mobility fleet,[6] and per-vehicle labels of faulty or abnormal vehicles in three releases of passenger EVs.[7–9] Two of those releases cover the same fleet and share 49 vehicles, 16 of them labeled abnormal, which are counted once.[8,9] The largest count in identified batteries, 491, is from a release of 1,027 lead-acid batteries of off-grid solar home systems that the operator diagnosed with capacity loss at repair.[29]

The largest count overall, 5,680, comes from a release of charging records from 30 public charging stations.[36] These records describe each charging session from the charger's side and carry no vehicle identifier. Batteries therefore cannot be followed over time as in the other releases, and each fault-labeled session is counted once. The release's README describes two kinds of fault: those reported by engineering personnel (such as thermal runaway, unbalanced cell voltages and short circuit) and over-temperature disconnections by the station. The labels do not say which kind applies to which fault.

The 30 faults reported by the authors are 28 battery systems returned to the manufacturer under warranty,[10] a vehicle pack with a defective cell that limits its performance[41] and a defective cell bypassed in a lead-acid string in grid service.[4]

**Table 3.** Faults in the collection, by release.

| Package Name | Faults | Batteries | Basis |
|---|---|---|---|
| shenzhen_charging | 5,680 | Not identified | Charging sessions labeled faulty (engineer-reported faults or over-temperature disconnection by the station, not distinguished). Counted per session, since there is no vehicle identifier. |
| aitio | 491 | 491 | Batteries that entered repair for capacity loss, diagnosed by the operator at repair (lead-acid). The authors selected a roughly balanced set (491 with capacity loss, 536 without), so this is not a fleet failure rate. |
| xie | 116 | 83 | Cells with a fault label (24 micro-short circuit, 28 low capacity, 35 high SOC, 29 low SOC) in 83 vehicles with a fault label. |
| cao | 61 | 61 | Vehicles labeled faulty. The paper names four fault types (thermal runaway, electrolyte leakage, internal short circuit and excessive aging). The release gives only faulty or not. |
| zhang2023 | 55 | 55 | Vehicles labeled abnormal. The release does not say what was abnormal. |
| evbattery | 32 | 32 | 48 vehicles labeled abnormal, less the 16 that are the same |

| Package Name | Faults | Batteries | Basis |
|---|---|---|---|
| | | | vehicles as in zhang2023. The release does not say what was abnormal. |
| schaeffer | 28 | 28 | All 28 systems returned to the manufacturer under warranty. |
| bilfinger2025 | 1 | 1 | A vehicle pack with a defective cell that limits its performance, reported by the authors. |
| m5bat-pbacid | 1 | 1 | The string with the defective cell bypassed on 10 August 2023. |
| **Total** | **6,465** | **752** | 785 faults in 752 batteries (755 in 722 from released labels, 30 reported by the authors), plus 5,680 fault-labeled charging sessions from one release without vehicle identifiers. |

## 3. Methods

### 3.1. Acquisition

Each release was downloaded from its original host (Table 1) and kept complete and unmodified, with archives left unextracted. The repository's fetch command downloaded 26 of the 35 releases: those on Zenodo, figshare, GitHub, Oxford ORA and the Stanford Digital Repository, plus five on Mendeley Data and mediaTUM. The command lists every file in the record and resumes interrupted downloads. Where the host publishes a checksum, it checks each file against it. For one of these releases, a utility-scale system,[5] most of the data (cell-level records) sit on a cloud drive and were downloaded through a web browser. The other 9 releases were downloaded from their host pages: three mediaTUM records whose download archive is rebuilt on each request, institutional servers, a manufacturer's portal and a cloud drive. Each release is held in its own folder with the data exactly as shipped and a short record of its source links, download date and contents.

### 3.2. Verification

Every held file was hashed: 7,205 files across the 35 releases. Where the host publishes a checksum (MD5, SHA-256 or SHA-512), our hash was compared with it. This covers 5,933 files and 109 archive members, and all of them match. For files without a published checksum, our own hash is recorded so that users can confirm they hold the same version we used.

### 3.3. Loaders

Each release has one loader in the repository. Measured counts and spans in this paper are derived by reading the data through these loaders. Every loader does two things: it lists the release's units (vehicles, storage systems or devices) and loads the data for a unit. For 30 of the 35 releases these are the same two functions, and the dashboard gives the exact call for each. Loaders read each deposit as published and return the released values unchanged, with the authors' column names where the release gives them. A schema note for each release lists its files, columns and units, and a notebook for each release documents its structure, signals and checks against the paper.

### 3.4. Terminology

Here we refer to a "unit" as one battery as defined in the corresponding dataset. This could be a vehicle, a stationary system, a portable system or a device. The rule for each release is recorded in the repository's counting rules, for example one archive per stationary system[11] or one file per vehicle.[24] Where a release reports cell counts or per-cell channels, that number is used as the cell count. Where the pack topology is stated, cells connected in parallel are each counted. For example, one release reports 108 cell-voltage channels, each for a pair of parallel cells, so its pack holds 216 cells (108 × 2).[34] Where no parallel count is stated, the reported number is used, so it may count series elements rather than cells. Where the release and its paper disagree, the release value is used in every total, and Table 2 gives the paper's figure in parentheses.

"Service time" is how long each unit was observed: the time from its first record to its last, summed over the units of a release. For one fleet[21] the per-vehicle spans come from a summary table that a script in the repository rebuilds from the raw archive. These spans come from relative timestamps, so they cannot be placed on a calendar. For one microgrid release[12] the span comes from the coverage stated by the authors. For one second-life storage release[35] the spans come from the installation dates in its module-info tables, clipped to the period of its exports. Eight other releases have no calendar timestamps. One of them, a summary table for four buses, has no time information at all. For the other seven, a lower bound in observation-hours is given instead:

1) Two releases ship short data snippets that can be used to calculate runtime.[8,9]
2) One fleet infers operating time using 30 s spacing between measurements, stated by the authors.[7]
3) One grid-scale release uses the duration stated by the authors.[26]
4) One satellite release sums the span of each data collection.[33]
5) One grid-storage release sums the recorded duration of every cycle of its field packs.[38]
6) One release of charging records counts the distinct minutes covered by its charging sessions.[36]

Chemistry labels are standardized to a short vocabulary: NMC, NCA, LFP, LMO, LMO/NMC (a blend), LCO, LTO, lead-acid, NiMH, vanadium flow and Li-ion. "NMC-class" marks releases where the chemistry is inferred from cell voltage. Such batteries could have NMC, NCA or LMO cathodes but cannot readily be told apart this way. "Li-ion" is used where the cathode chemistry is neither stated nor inferable from the data. Lead-acid includes flooded, gel and valve-regulated designs. One release of charging records also names lithium polymer, which is a cell format rather than a chemistry, so it is listed as stated.[36] In the tables, SOC and SOH stand for state of charge and state of health as estimated by the BMS.

### 3.5. Energy Measurement

Energy is reported for 28 of the 35 releases: about 115.6 MWh in total, of which 35.84 MWh is a rating stated in the paper or data record (basis: stated) and 79.76 MWh is computed (Table 4). For four releases the energy covers only some of the units: 3 of 28 vehicles,[30] 1 of 4 households,[31] 473 of 514 vehicles[7] and 9 of 11 grid-scale systems, not counting the one shared with another release.[39] These are marked § in Table 2. Where a rating is not stated, energy is computed (basis: computed) in one of two ways:

1) From the stated configuration, it is the number of cells or modules times their rated capacity and nominal voltage, using the release's counts and the paper's ratings.
2) From the released data, in one of three ways:
    a. Coulomb counting: the charge in ampere-hours over a long charge or discharge, divided by the change in state of charge, times the mean pack voltage over the same period, taking the median per battery.
    b. Energy counter: the vehicle's own cumulative energy counter divided by the change in state of charge.
    c. Consumer devices: each device's reported nominal capacity times its median cell voltage.

Values computed from the data refer to usable energy, not nameplate ratings, with two exceptions: one release's value is divided by the BMS state of health, so it approximates the rated energy,[15] and the consumer-device values are nominal.[23] One release is part of a larger plant report that also covers a lead-acid string released separately.[4,20] That string's energy is counted once under its own release. One grid-scale release normalizes power and current to ratings it does not release, so its energy is each system's documented rating.[39,44] For one system, the documentation gives 10 MW and 30 MWh, but the released peak power of 0.25 C and the documented energy-to-power ratio of 4 both imply 40 MWh, which is used (basis: computed).

**Table 4.** Energy per release and how it was established, in the same order as Table 1. Stated: a rating from the paper or record. Computed: from the stated cell configuration or from the released data.

| Package Name | Energy, kWh | Basis | Method |
|---|---|---|---|
| tsukuba | 326.0 | Stated | Rating stated in paper |
| ved | 72.0 | Stated | Rating stated in README (3 EVs × 24 kWh), PHEVs unrated |
| brazil_star | 1.2 | Computed | Coulomb counting over long SOC swings, one of four households |
| aitio | 246.5 | Computed | Configuration: 1,027 × 12 V × 20 Ah |

| Package Name | Energy, kWh | Basis | Method |
|---|---|---|---|
| beijing_ev | 32.0 | Computed | Coulomb counting over long SOC swings |
| birds_cubesat | 0.06 | Computed | Configuration: 4 × 6 NiMH cells × 2.0 Ah × 1.2 V |
| cloverleaf | 79.7 | Computed | Coulomb counting over the typical-day sheet, divided by SOH so it approximates the rated energy, times 3 packs |
| deng | 952.6 | Computed | Configuration: 20 × 90 cells × 145 Ah × 3.65 V |
| bilfinger2024 | 110.5 | Stated | Rating stated in paper |
| m5bat-2023-04 | 6,714.0 | Stated | Rating stated in report, less the unit released separately |
| rwth-home | 152.6 | Stated | Ratings in released system metadata |
| schaeffer | 107.5 | Computed | Configuration: 28 × 24 V × 160 Ah |
| bilfinger2025 | 372.6 | Stated | Net energy stated in paper for the five vehicles not counted elsewhere (58, 2 × 82.3, 2 × 75 kWh) |
| cao | 15,107.0 | Computed | Coulomb counting at the stated 30 s spacing, times summed cell voltage, median per vehicle by brand, one brand excluded |
| changan | 16,293.6 | Computed | Configuration: 300 × 96 cells × 155 Ah × 3.65 V |
| ku_leuven_bev | 139.3 | Stated | Ratings stated in paper (60.5 and 78.8 kWh) |
| rise_conbat | 115.1 | Stated | Pack ratings in release metadata (9.9 or 13.0 kWh) divided by 8 modules per pack |
| tongji_rest | 188.7 | Computed | Coulomb counting over long SOC swings, 4 field vehicles |
| tumftm | 406.0 | Stated | Rating stated in paper (58 kWh net per vehicle) |
| bilfinger2026 | 400.5 | Stated | Rating stated in paper |
| cell_to_pack | 6,500.5 | Computed | Coulomb counting over long SOC swings, 16 field packs |
| li2026 | 156.0 | Computed | Configuration: 240 cells x 200 Ah x 3.25 V (cell rating from paper) |
| m5bat-pbacid | 1,066.0 | Stated | Rating stated in paper |
| pnnl_epri | 63,870.0 | Stated, computed | Documented ratings of 8 systems, one system computed from the data (40 MWh), 2 systems unrated, and the ppl system counted under ppl |
| ppl | 2,000.0 | Stated | Rating stated in paper |
| rwth-android | 0.7 | Computed | Device nominal capacity times median cell voltage (nominal, not measured usable energy) |
| stanford_ev | 91.4 | Computed | Coulomb counting over long SOC swings |
| zhou2026 | 209.0 | Stated | Nominal pack energies stated in the supplementary information (2 x 52 kWh cars, 105 kWh bus) |
| **Total** | **115,601** | | 28 releases. The two vehicles shared by bilfinger2024 and bilfinger2026 counted once (-110.5 kWh), and the ppl system in pnnl_epri counted once, under ppl |

### 3.6. Faults and Shared Units

Faults are counted once per faulty cell where a release labels individual cells, and once per battery otherwise, so a battery with several labeled cells can carry several faults (Table 3). One release has no vehicle identifier, so each of its fault-labeled charging sessions is counted once and the batteries involved are not identified. Only labels released with the data and faults the authors report in the paper are counted. Temporary outages, measurement artifacts, deliberate test events and laboratory-emulated faults are not counted.

Six overlaps between releases are counted once, under one home release (Table 5). Their cells are also counted once, which removes 538 cells. The releases number their units differently, so shared units were found by matching their data. If another release observed a shared unit over a period its home release did not cover, that time is added to service time.

**Table 5.** Units that appear in more than one release, each counted once.

| Releases | Shared units | How identified | How counted |
| --- | --- | --- | --- |
| evbattery and zhang2023 | 49 vehicles | Identical data snippets and agreeing fault labels | Under zhang2023, which holds more of their data and their fault labels |
| bilfinger2024 and bilfinger2026 | 2 vehicles | The same vehicles measured again | Under bilfinger2024, later charges count |
| m5bat-2023-04 and m5bat-pbacid | 1 lead-acid string | Its voltage record | Under m5bat-pbacid |
| ppl and pnnl_epri | 1 grid-scale system | Matching state of charge and cell voltage | Under ppl, 304 earlier days (0.83 unit-years) count |
| tumftm and tum_id3_2022 | 1 vehicle | Identical state of charge and pack voltage | Under tumftm with its 216 cells, adds pack current and 108 cell voltages |
| bilfinger2024, bilfinger2026 and bilfinger2025 | 7 vehicles | Identical earlier recordings and license plates | Under bilfinger2024 and bilfinger2026, later charges count |

## 4. Recommendations for Future Releases

Assembling these 35 releases showed what makes field data easy or hard to reuse. On the battery side, the hardest gaps were in identity and context. Two releases of the same fleet[8,9] number the same vehicles differently, so shared vehicles could only be found by matching their data. A relative counter[8,9,21] or no timestamps at all[7,19] rules out calendar aging, seasonality and links to weather. One release[7] mixes measured cell voltages with unlabeled outputs of the authors' own model, which took the supplementary information and the authors' code to untangle. Missing pack topology, ratings and sign conventions make releases hard to compare.

On the data-handling side, two releases state no license for their data. One release presents 133 vehicles and publishes three.[27] One grid-scale release drops signals from 2022 onward without saying why.[5] Placeholder values can look like faults unless documented, as one lead-acid codebook does.[4] As of October 2026, users of this collection download 9 of the 35 releases themselves. Three releases use RAR archives, and one is a 79.7 GB split archive. Table 6 turns these recommendations into a checklist for future releases.

**Table 6.** Checklist for publishers of battery field data. Unmarked items are needed for some uses of the data. Items marked ‡ make reuse easier.

| Area | Recommendation |
|---|---|
| **Battery-specific** | Give every battery an ID, and use the same IDs across files and releases. |
| | Record calendar timestamps, not relative counters. |
| | Release per-cell voltages where you have them. |
| | Say what each fault label refers to and how it was decided. |
| | Mark BMS estimates (state of charge, state of health) as estimates. |
| | State the pack topology, nominal capacity and voltage. |
| | State the sign convention for current and power.‡ |
| **Data hygiene** | State the license on the data record. |
| | Name every column and give its unit. |
| | Keep raw data separate from processed or model outputs, and describe any processing. |
| | Deposit data in a repository with a download interface, such as Zenodo or figshare.‡ |
| | List the placeholder values used for missing readings. |
| | State which part of the study the release covers. |
| | Document any change in the recorded signals over time. |
| | Check that exports are complete, and publish file checksums.‡ |
| | Use open file and archive formats (such as CSV, parquet and zip), not pickled Python objects or RAR, and avoid split archives.‡ |

## 5. Usage Notes

Getting started takes five steps. The dashboard (docs/dashboard, published at https://astrolabe-analytics.github.io/public-battery-field-data/dashboard/) lists every release with its download route, the code to load it and a preview plot.

1. **Install** the repository (or open the quick-start notebook in Google Colab, no install needed).
2. **Set your data folder,** where releases will be saved.
3. **Fetch a release.** As of October 2026:
   - 25 releases download with one command.
   - For 9 releases, the command lists which files to download from the host page and where to save them.
   - One utility-scale system downloads its main files with one command, and its cell-level data from a cloud drive the command links to.

   Start with a release under 1 GB. The full collection is 256 GB, mostly in four releases (Table 2).
4. **Check the setup** with `python -m fielddata.doctor`, which reports any missing, misnamed or incomplete release folder in plain language.

5. **Run the quick-start notebook.** Set one line to a release name, and it lists the units, loads a slice of one unit and plots a signal. Most releases then load with the same two functions. The dashboard shows the exact call for each release.

- **Per-release notebooks** show each release's file structure, signals and checks against its paper. The 514-vehicle fleet[7] has an in-depth worked example.
- **Columns, units and placeholder values differ between releases. Each release's schema note documents them, and nothing** is renamed or converted by default.
- **Three releases use RAR archives**, which need an unpacking tool such as unrar or libarchive.
- **License terms differ.** Six releases restrict commercial use, one of them also forbidding derivatives, and two state no license for their data (Table 1). Check each release's terms before reuse. The release that forbids derivatives[40] is read from its original files only, with no preview plot and no derived data in the repository.

### Data Availability

Each release is available from its original record, linked in Table 1. The repository, including the registry of sources, checksums, loaders, tables and notebooks, is archived on Zenodo (https://doi.org/10.5281/zenodo.23150592) and developed at https://github.com/Astrolabe-Analytics/public-battery-field-data. To report an error or suggest a dataset we missed, please open an issue on the repository.

### Code Availability

All code is in the repository under the MIT license. Every headline number in this paper is produced by one command (python scripts/rebuild.py) from input files stored in the repository. Other numbers cite a row of the repository's package facts table or a named source.